\documentclass[final,authoryear,3p,times,twocolumn]{elsarticle}

 \usepackage{graphicx}
 \usepackage{ulem}

\usepackage{amssymb}
\usepackage{multirow}
\usepackage{array}
\usepackage{rotating}

\newcommand{\hartley}{103P/Hartley~2}
\newcommand{\halley}{1P/Halley}
\newcommand{\cam}{a$^3\Pi$}

\journal{Planetary and Space Science}

\begin{document}
\begin{frontmatter}



\title{\bf  Model for the Production of CO Cameron band emission in Comet 1P/Halley }


\author{Susarla Raghuram and Anil Bhardwaj \corref{cor1}}
\cortext[cor1]{Corresponding author: anil\_bhardwaj@vssc.gov.in, bhardwaj\_spl@yahoo.com}
\address{Space Physics Laboratory, 
Vikram Sarabhai Space Centre, Trivandrum 695022, India.}

\begin{abstract}
The abundance of CO$_2$ in comets has been derived using CO Cameron band 
(a$^3\Pi\rightarrow$ X$^1\Sigma^+$) emission
assuming that photodissociative excitation of CO$_2$ is the main production process of CO(\cam).
On comet \halley\ the Cameron (1-0) band has been observed by International Ultraviolet Explorer (IUE)
on several days in March 1986.  A coupled chemistry-emission model is developed for comet 
\halley\  to assess the importance of various production and loss mechanisms of CO(\cam) 
and  to calculate the intensity of Cameron band emission on different days of IUE observation.  
Two different solar EUV flux models, EUVAC of \cite{Richards94} and  SOLAR2000 of \cite{Tobiska04},
and different relative abundances of CO and CO$_2$, are used to evaluate the role of photon 
and photoelectron in producing CO molecule in \cam\ state
 in the cometary coma. It is found that  in comet \halley\ 60--70\%  of the total intensity 
of the Cameron band emission is contributed  
 by electron impact excitation of CO and CO$_2$, while the contribution from  photodissociative 
 excitation of CO$_2$ is small (20--30\%). Thus,
 in the comets where CO and CO$_2$ relative abundances are comparable, the 
Cameron band emission is largely governed by electron impact excitation of CO,
 and not by the photodissociative  excitation of CO$_2$ as assumed earlier. 
Model calculated  Cameron band  1-0 emission intensity (40 R) is consistent with the observed  
IUE slit-averaged brightness (37 $\pm$ 6 R) using EUVAC model solar flux on 13 March 1986, and also 
on other days of observations.
Since electron 
impact excitation is the major production mechanism, the Cameron  emission can be used to 
derive photoelectron density in the inner coma rather than the CO$_2$ abundance. 

\end{abstract}


\end{frontmatter}


\section{Introduction}
Ejecting neutral gas and dust into space, comets  create extensive and  unique atmospheres 
in the interplanetary space. Interaction of solar extreme ultraviolet (EUV) radiation with cometary 
species causes spectrum of different  emissions.
 Spectroscopic observations of comets   in the UV region
by space-based telescopes give information about composition, 
abundance, and spatial distribution of neutral species in the cometary coma \citep[e.g.,][]{Feldman04}.
The number densities of  CO$_2$ and CO in cometary coma have been derived 
 using emissions from the dissociative products which can be produced in  
 metastable states.  Assuming photodissociative excitation is the main
 production mechanism in populating the a$^3\Pi$ metastable state of CO, the Cameron band
 (a$^3\Pi$ $\rightarrow$ X$^1\Sigma^+$) emission has been used to estimate 
  the abundance of CO$_2$ in comets \citep{Weaver94,Weaver97,Feldman97}.

 The observation of Cameron band of CO molecule in the coma of 
 comet \hartley\ \citep{Weaver94} by Hubble Space Telescope (HST) gave an incitement to re-examine 
 the data of several comets observed by the International Ultraviolet Explorer (IUE)
 satellite.   Cameron band (1-0) emission at 1993~\AA\ is observed in 4 comets, including comet \halley, 
 in the IUE spectra \citep{Feldman97}. The Cameron band  (0-0) and (0-1) emissions at 2063 and 
 2155~\AA, respectively, could not be observed since they fall in the low sensitivity
  end of the IUE long-wavelength camera.   Since the excited upper state (\cam) of 
 Cameron band emission  is  metastable and its lifetime 
  is very small \citep[$\sim$3 ms,][]{Gilijamse07} compared to lifetime of CO$_2$ molecule 
 \citep[$\sim$135 hours at 1 AU,][]{Huebner92}, 
  the CO(\cam) molecule can travel a distance of few meters only in the cometary coma  
  before de-exciting into ground state (X$^1\Sigma^+$)  via emitting photons. Hence, the Cameron band 
  emission can be used to probe CO$_2$ distribution, and thus its abundance in the coma, provided 
  it is produced only through photodissociation of CO$_2$.

Besides photons, the solar EUV-generated photoelectrons also play a significant
role in driving the chemistry of cometary species in the coma. 
The importance of photoelectrons in excitation, dissociation, and ionization  
of various cometary species and subsequent effects on emissions in the inner
coma are discussed in several works 
\citep[e.g.,][]{Cravens78,Ip85,Boice86,Korosmezey87,Bhardwaj90,Bhardwaj96,Haider93,Haberli96,
Bhardwaj99a,Bhardwaj03,Haider05,Campbell09,Feldman09,Bhardwaj11a,Bhardwaj11b}. 
 To explain the Cameron band emission in comet \hartley, \cite{Weaver94} considered five possible 
 production mechanisms of CO(\cam) molecule. The modelled Cameron 
 band emission of CO molecule  by \cite{Weaver94} suggested that 60\% of 
 total  CO(\cam) production  can be through photodissociative excitation of CO$_2$; the remaining 
   was attributed  to other excitation processes. \cite{Feldman97} assumed that photodissociative 
 excitation of CO$_2$ is the only source of Cameron band emission in comet \halley. 
 Recent calculations of \cite{Bhardwaj11a} have
 demonstrated that in the comet \hartley, 60 to 90\% of CO(\cam) production is through the photoelectron
 impact of CO$_2$ and  CO and that the contribution of photodissociation  of CO$_2$ 
 is quite small.  
The derived rates of electron impact dissociation of CO$_2$ producing CO(\cam) by
\cite{Feldman09}  show that photodissociation can be  comparable with electron impact excitation 
in producing Cameron band emission.
 However, the comet \hartley\ is depleted in CO  (relative abundance $<$1\%). But in the
 case of comet \halley\ the CO abundance is relatively higher compared to that on the  \hartley,
 and hence the contribution due to direct excitation of CO by electron impact would be much larger.

 There are  several  observations of CO in comet \halley, as well as in other comets,
 which suggest that CO is produced directly from the nucleus 
 as well as having prevailed distributed sources in the cometary coma 
\citep[]{Eberhardt87,Eberhardt99,Disanti03,Cottin08}.
The measured number density of CO by neutral mass spectrometer on Giotto spacecraft, which flew 
through the coma of \halley, is $\le$7\% relative to water at 1000 km cometocentric distance. 
This relative abundance is  higher ($\le$15\%) at larger distances (2 $\times$ 10$^{4}$ km) 
in the  coma \citep{Eberhardt87,Eberhardt99,Festou99}. This increase in abundance can be explained
 by dissociation of CO-bounded species and also through heating of several refractory grains by sunlight. 
Other cometary species like H$_2$CO, C$_3$O$_2$,  POM (polyoxymethylene, or polyformaldehyde),
 CH$_3$OH, and CO$_2$ can also produce CO molecules in photodissociation process 
\citep[see][and references there in]{Greenberg98,Cottin08}. However, there are no literature reports 
on the production of CO(\cam) from CO-bearing species, like H$_2$CO, CH$_3$OH, and C$_3$O$_2$, via 
photodissociation or electron impact dissociative excitation.

 Reanalysis of the IUE data on comet \halley\ showed  5 observations of the Cameron 1-0 band emission, which  
 span over a 10-days period in March 1986;  the intensity of 1-0 emission  varied by a factor of about 4 from lowest 
value of 20$\pm$6 to highest value of  65$\pm$9 Rayleighs \citep{Feldman97}. 
 Assuming that the production of Cameron band emission is only through photodissociation of CO$_2$, 
 \cite{Feldman97} derived the CO$_2$ abundances of $\sim$2 to 6\%, and also the CO$_2$/CO abundance ratio.

The production of CO(\cam) is
mainly associated with spatial distribution of CO$_2$ and CO molecules in the coma. 
We have recently developed a model for the chemistry of CO(\cam) on  comet \hartley\
\citep{Bhardwaj11a}. In the present paper this coupled chemistry model has been employed to 
study the production of Cameron band emissions on comet \halley.
The contributions of  major production and loss processes  of CO(\cam) in  comet \halley\ are 
evaluated for different relative abundances of CO and CO$_2$.

The photochemistry in the cometary coma is driven by  solar UV-EUV radiation.
The solar UV flux is  known to vary considerably both with the 27-day solar rotation period and 
with the 11-year solar activity cycle. Since the continuous  measurements of solar EUV fluxes are not 
available for different cometary observations, one has to depend on the empirical solar EUV models.
 To assess the  impact of solar EUV flux on the calculated brightness
of Cameron band  emission we have taken two most commonly used 
solar EUV flux models, namely EUVAC model of \cite{Richards94} and SOLAR2000 v.2.3.6 (S2K) 
model of \cite{Tobiska04}. The solar EUV flux from these two models for 13 March 1986 
are shown in Fig.~\ref{solflx}.

This paper will demonstrate that in comets where 
CO$_2$ and CO relative abundances are comparable, 
the photoelectron impact excitation of CO plays a major role in controlling the brightness of Cameron 
band, and not the photodissociation of CO$_2$ as assumed previously. 
Since the Cameron band  emission is forbidden 
and electron impact is the major excitation mechanism, this emission is suitable to track photoelectron 
flux in the inner cometary coma rather than the CO$_2$ abundance. 
We have also studied  the sensitivity of calculations associated with variation in input 
solar flux and  electron impact excitation cross sections of CO$_2$ and CO in estimating 
the intensity of Cameron band  emission.

\section{Model}
The neutral parent species considered in the model are H$_2$O, CO$_2$, and CO. 
The density of neutral parent species in the coma  is calculated 
using Haser's formula, which assumes spherical distribution of gaseous environment 
around the nucleus. The number density n$_i(r)$ of i$^{th}$ species in the 
 coma at a cometocentric distance $r$  is given by 
\begin{equation}
 n_i(r)=\frac{f_iQ_p}{4\pi v_ir^2} (e^{-\beta_i/r})
 \label{haser}
 \end{equation}
Here Q$_p$ is the total gas production rate of the comet, $v_i$ is the average velocity of
neutral species taken as 1 km s$^{-1}$, $\beta_i$ is scale length ($\beta_{H_2O}$
= 8.2 $\times$ 10$^{4}$ km, $\beta_{CO_2}$ = 5.0 $\times$ 10$^{5}$
km, and $\beta_{CO}$ = 1.4 $\times$ 10$^{6}$ km) and f$_i$ is fractional
abundance of i$^{th}$ species. Calculations are made for comet 1P/Halley taking total gas 
production rate as 6.9 $\times$ 10$^{29}$ s$^{-1}$, which has been 
observed by Giotto  mission \citep{Krankowsky86}. Since cometary coma is dominated by water, 80\%
of total production rate is assumed to be H$_2$O. 

 The in situ gas measurements  at comet \halley\ 
made by  Giotto Neutral Mass Spectrometer~(NMS)  on the encounter date 13 March 1986 
showed that  CO$_2$ abundance is 3.5\% of water \citep{Krankowsky86}. On the same day, 
based on IUE observation, \cite{Feldman97} derived CO$_2$ abundance of 4.3\%.
 \cite{Eberhardt87} suggested that below 1000 km,  nuclear rate of  CO production 
 can be 7\% of water. The radial profile of CO calculated by \cite{Eberhardt87} showed  almost 
a constant value of CO relative abundance ($\le$15\%) above 15000 km.
 This increase in CO abundance is  attributed to the presence of an extended source for CO in the cometary coma.
The IUE-derived average production rate of CO is 4.7\% \citep{Feldman97}. We have taken 4\% CO$_2$ and 
7\% CO  directly coming from nucleus as  the standard input for the model.
We have also considered extended CO density profile directly from Giotto NMS observation
 \citep{Eberhardt87}.  Further, the relative abundances of CO$_2$ 
and CO are varied to assess the effect on the intensity 
of Cameron band emission and different production channels of CO(\cam).

The primary photoelectron energy spectrum  $Q(E,r,\theta$), at energy E, cometocentric distance r,
and solar zenith angle $\theta$, is calculated by degrading the solar UV-EUV radiation
in the cometary coma using the following equation
\begin{equation}
Q(E,r,\theta) = \sum_{i}\int_{\lambda}n_i(r)\ \sigma_i^I(\lambda)\ I_{\infty}{(\lambda)}\
exp[-\tau(r,\theta,\lambda)]\ d\lambda 
\label{pheprod}
\end{equation}
where,
\begin{equation}
\tau(r,\theta,\lambda)= \sum_{i}\sigma_i^A(\lambda)\ sec\theta \int_r^\infty  n_i(r') dr'  
\end{equation}

Here $\sigma_i^A(\lambda)$  and $\sigma_i^I(\lambda)$ are absorption and ionization 
cross sections, respectively, of i$^{th}$ species at wavelength $\lambda$, 
n$_i$(r) is its neutral gas density calculated using the equation~\ref{haser}, and
 $\tau(r,\theta,\lambda)$ is  optical depth of the medium.
 $I_{\infty}(\lambda)$ is unattenuated solar  flux at the top of atmosphere at wavelength $\lambda$. 
 All calculations are made at solar zenith angle 0$^0$.
The photoabsorption and photoionization cross sections of H$_2$O, CO$_2$,
and CO are taken from \cite{Schunk09}.

The steady state photoelectron fluxes are calculated using the Analytical 
Yield Spectrum (AYS)
approach, which is based on the Monte Carlo method. Details of the AYS approach are
given in several of the previous  papers \citep{Singhal84,Bhardwaj90,Bhardwaj96,Singhal91,
Bhardwaj93,Bhardwaj99a,Bhardwaj03,Bhardwaj99d, Bhardwaj99b,Haider05,Bhardwaj09}. 
We have used two dimensional yield spectra to calculated photoelectron flux $f_p(E,r)$ 
as a function of energy E and cometocentric distance r 
\begin{equation}
 f_p(E,r)=\int_{w}^{\infty} \frac{Q(E,r,\theta)U^c(E,E_o)}{\sum n_i(r)\sigma_{iT}(E)}
\end{equation}
where $Q(E,r,\theta)$ is primary photoelectron production rate calculated using equation~\ref{pheprod}.
$\sigma_{iT}(E)$ is  total inelastic electron impact cross section at energy E for the $i^{th}$ species whose
  number density is $n_i(r)$. The lower limit of integration $w$ is minimum excitation energy 
and $U^c(E,E_o)$ is two dimensional 
composite yield spectra \citep{Singhal84,Bhardwaj90}. The total inelastic electron impact cross sections 
 for water are taken from \cite{Rao95}, and those for CO$_2$ and CO are taken from \cite{Jackman77}.
 
The loss process of photoelectrons
through collisions with thermal electrons is  considered using the following formula
\begin{equation} 
n_e\sigma_{e-e}^{eff}=\frac{n_e\beta(E,n_e,T)}{E\ \overline{W}}
\end{equation}
where n$_e$ is the thermal electron density, E is the energy of photoelectron,
and $\overline{W}$ is the average energy lost per collision between
photoelectron and the thermal electron. The expression $\beta$ is given by \cite{Mccormick76}.
More details are provided in \cite{Bhardwaj90}. The calculated photoelectron fluxes for the two solar
 EUV flux models  at 1000 km are shown in Fig.~\ref{csca3pi}.

The electron impact volume  production rates of different ions from neutral species and volume excitation rates 
for CO(\cam) state from CO$_2$ and CO are calculated using photoelectron flux $f_p(E,r)$ and electron 
impact excitation cross section $\sigma_{ik}$ of i$^{th}$ species and k$^{th}$ state as
\begin{equation}
 V(r)=n_i(r)\int_w^{100}f_p(E,r)\ \sigma_{ik}(E)dE
\label{volex}
\end{equation}
 The cross sections for electron impact dissociative ionization  of  water
are taken from \cite{Itikawah2o}, for CO$_2$ from \cite{Bhardwaj09}, and for CO
 from \cite{Mcconkey08}.  

Table~\ref{taba3pi} presents the reactions involved in 
the production and loss of CO(\cam). \cite{Huebner92} calculated the cross section for 
photodissociative excitation of CO$_2$ producing CO in a$^3\Pi$ state using  total absorption 
cross section and the yield measured by \cite{Lawrence72}. We averaged these cross section values 
over 50 \AA\ bin intervals to calculate photodissociative excitation rate using solar flux from EUVAC and 
S2K models; 
this cross section is shown in Fig.~\ref{phota3pi}.
The cross section for electron impact excitation of CO in the a$^3\Pi$ state is taken from
\cite{Jackman77} and for dissociative excitation of CO$_2$ producing CO(\cam) is taken from
\cite{Bhardwaj09}. These cross sections are presented in Fig.~\ref{csca3pi}. 
To estimate the effect of electron impact cross sections on emissions, we have used 
the electron impact cross sections recommended by \cite{Avakyan98}  for the above two processes, which
are also shown in Fig.~\ref{csca3pi}. The electron temperature profile, required 
for dissociative recombination reactions, is taken from \cite{Korosmezey87}.

\section{Results and discussions}
\subsection{Cameron band emission}
The first clear observation of Cameron band emission of CO molecule is made  
in comets \hartley\ and C/1992 T2 Shoemaker-Levy \citep{Weaver94}, which was followed by detection 
in several other comets, including  1P/Halley, in the IUE reprocessed data 
\citep{Feldman97}. Assuming that the photodissociative excitation  of CO$_2$ is the major
production mechanism of Cameron band emission, \cite{Weaver94} derived  the  abundance of 
CO$_2$ in comet \hartley. Recently, \cite{Bhardwaj11a} have   demonstrated that on comet \hartley\
 the photoelectron impact dissociative excitation of CO$_2$ followed by photoelectron
impact of CO are the  major production processes of Cameron band, and not
the photodissociative excitation of CO$_2$ as suggested by \cite{Weaver94}.

Since comet \hartley\ is  CO  depleted (relative abundance $\le$1\%), 
the contribution to Cameron band emission through dissociative
excitation of CO$_2$ by   EUV-generated photoelectrons is more important. However, in case of
comets where CO abundance is larger, like \halley, the contribution of CO to the Cameron band
 emission  would be significant. The derived CO$_2$/CO abundance ratios for several IUE observations
of comet \halley\  showed that the abundance of CO can be even double that
 of CO$_2$ \citep{Feldman97}.

\begin{figure}  
\begin{center}
\noindent\includegraphics[width=20pc]{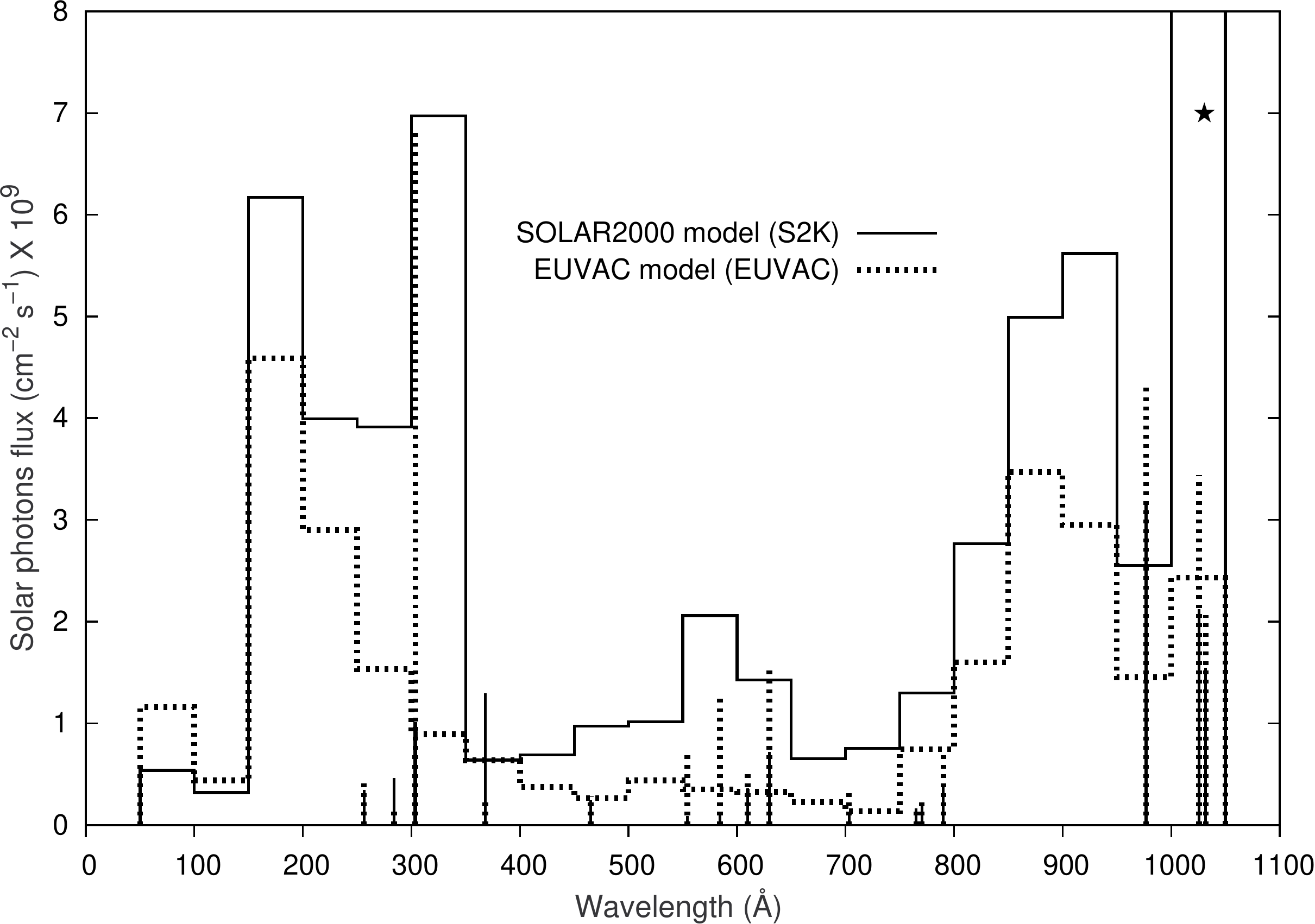}
\caption{Solar EUV fluxes from EUVAC model \citep{Richards94} and SOLAR2000 (S2K) model
\citep{Tobiska04} for the day 13 March 1986. Significant differences in the two model solar
EUV fluxes can be noticed. \small({\ding{72}})~The value of solar flux  in SOLAR2000 model 
for the bin 1000--1050~\AA\ is 30 $\times$ 10$^{9}$ cm$^{-2}$ s$^{-1}$.}
\label{solflx}
\end{center}
\end{figure}

\begin{figure}  
\begin{center}
\noindent\includegraphics[width=20pc,angle=0]{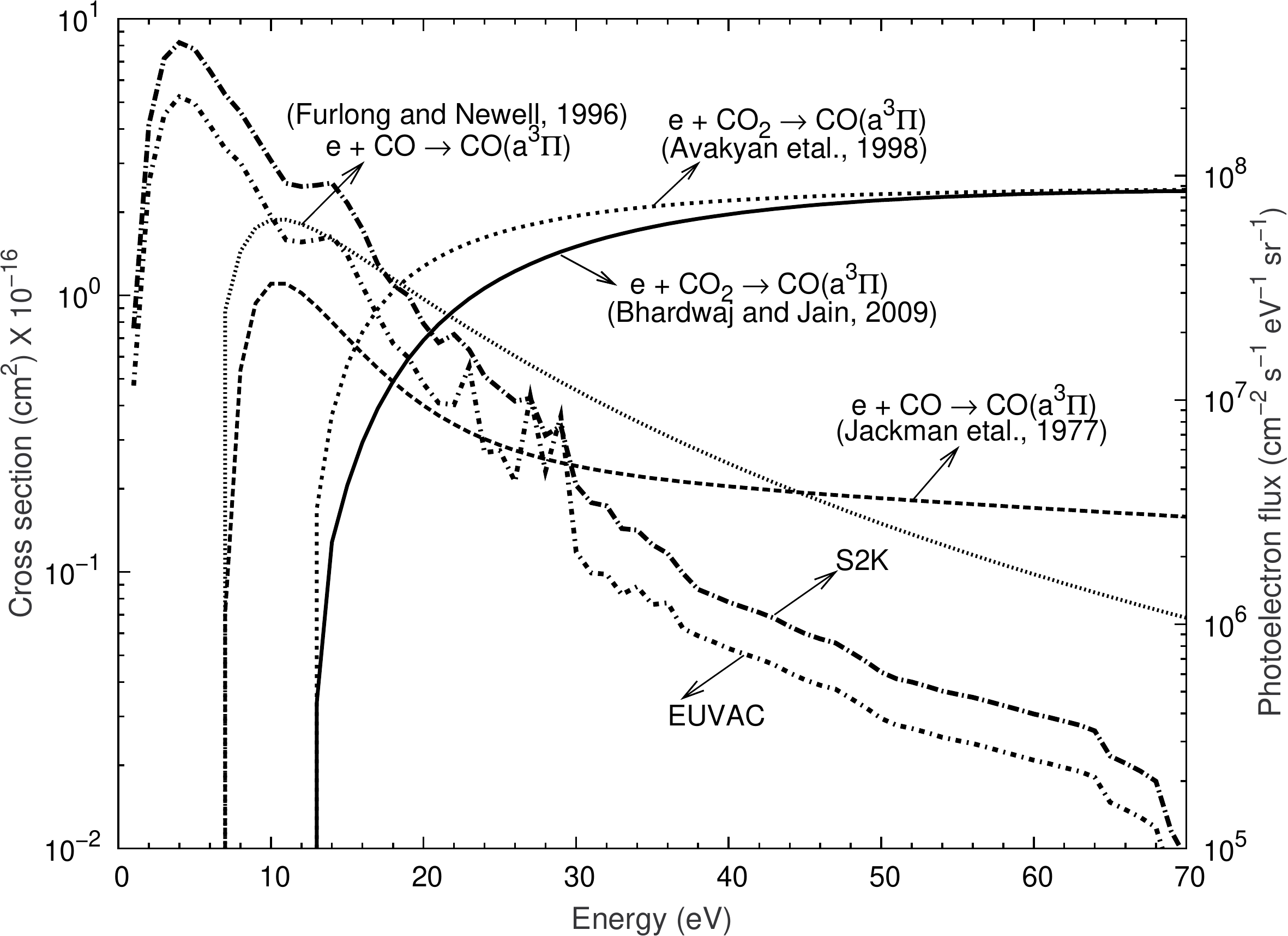}
\caption{Cross sections for electron impact excitation  of CO(\cam)
 from CO and CO$_2$. Calculated photoelectron flux at cometocentric
 distance of 1000 km  is also shown for both SOLAR2000 (S2K) and EUVAC model solar fluxes
with magnitude on right side y-axis}.
\label{csca3pi}
\end{center}
\end{figure}

\begin{figure}  
\begin{center}
\noindent\includegraphics[width=20pc,angle=0]{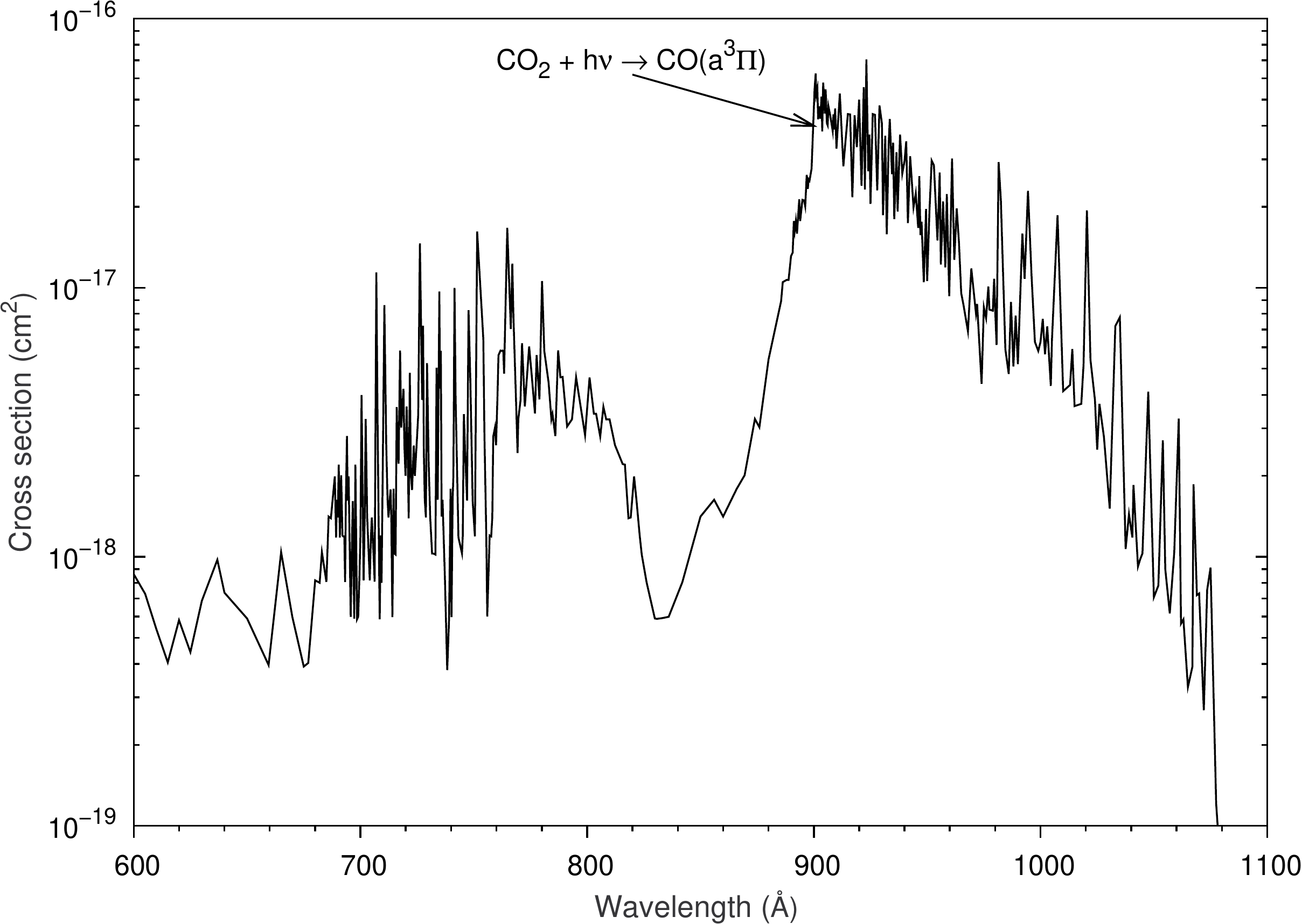}
\caption{Photodissociative  excitation cross section of CO$_2$ producing CO(\cam),
 taken from \cite{Huebner92}. }
\label{phota3pi}
\end{center}
\end{figure}

\begin{figure}  
\begin{center}
\noindent\includegraphics[width=20pc,angle=0]{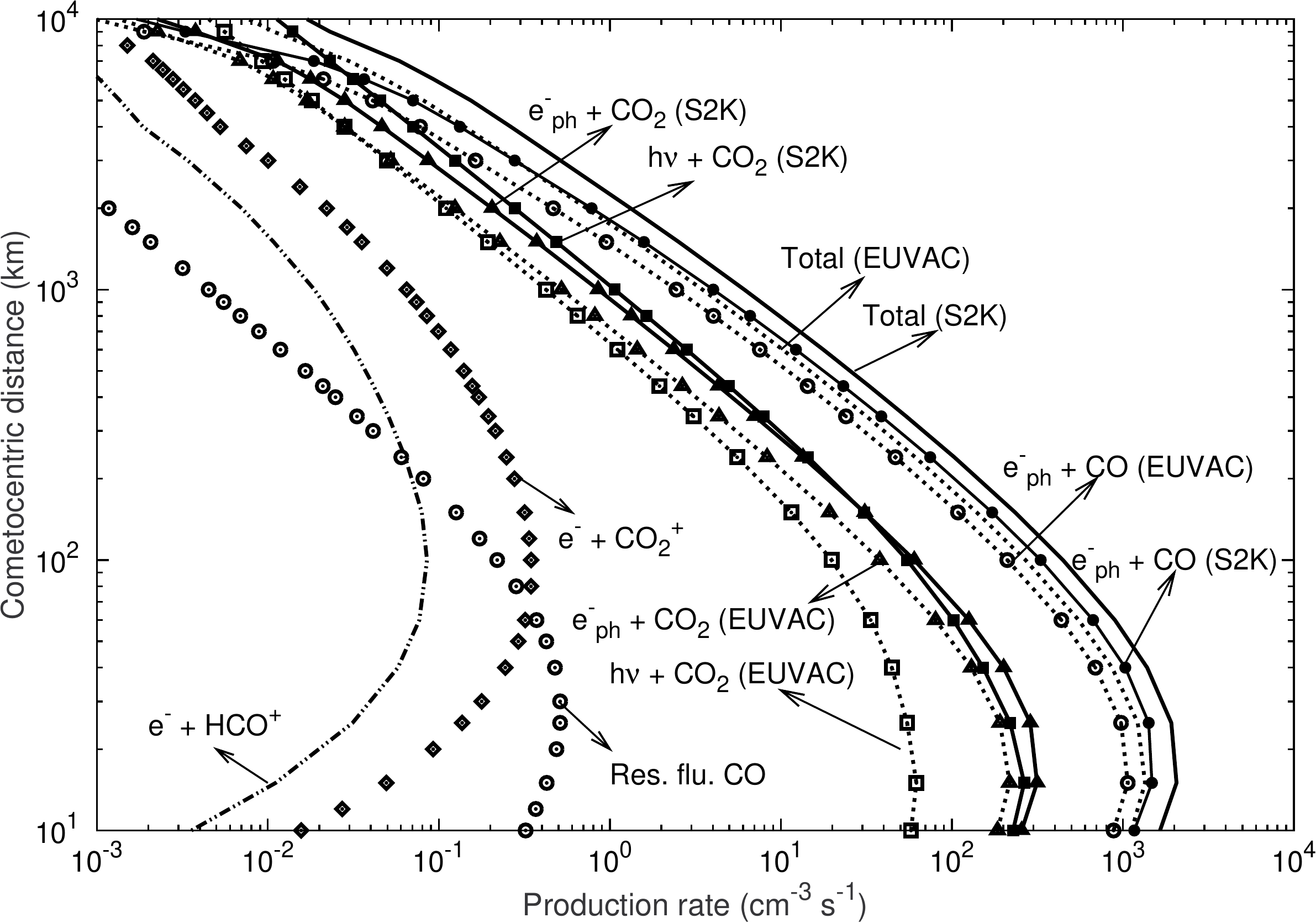}
\caption{Radial profiles of various production mechanisms of CO(a$^3\Pi$) in comet \halley\ on 13
March 1986 for relative abundance of 4\%  CO$_2$ and 7\% CO. 
The calculated profiles for dissociative recombination of CO$_2^+$ and HCO$^+$, and resonance 
fluorescence of CO are shown for EUVAC solar flux only. Res. flu. = resonance fluorescence of CO molecule. 
e$^-_{ph}$ = Photoelectron, h$\nu$ = Solar photon, and e$^-$ = thermal electron.}
\label{proda3pi}
\end{center}
\end{figure}

The calculated production rate profiles of CO(\cam)  using solar EUVAC and S2K  models 
 for relative abundance of 4\% CO$_2$  and 7\%   CO 
are shown in Figure~\ref{proda3pi}. For  both solar EUV flux models,  
the peak production rate  occurs at cometocentric distance $\sim$20 km.
The major production mechanism of CO(\cam) is the photoelectron impact of CO, whose contribution 
is $\sim$70\% to the total CO(\cam) production. On using the S2K solar flux, the calculated  
total production rate is  1.5 times larger  than that obtained using the EUVAC  flux.  
 This variation is mainly due to the difference  in the input solar EUV flux (cf.~Fig.~\ref{solflx})
and subsequently  EUV-generated photoelectron flux (cf.~Fig.~\ref{csca3pi}).
In the  wavelength region 700--1050 \AA, the  S2K model solar flux is a factor of 
$\sim$2.5 larger than the EUVAC model (cf.~Fig.~\ref{solflx}).
 As shown in Figure~\ref{phota3pi}, the photodissociative 
excitation cross section of CO$_2$ producing CO(a$^3\Pi$) maximizes around  
880--1000 \AA. Further, the S2K solar flux in the 1000--1050~\AA\ wavelength bin is around 20 times 
higher than the EUVAC flux. The average cross section 
value for photodissociation of CO$_2$ producing CO(\cam)  in the wavelength  region 1000--1050~\AA\ 
is  comparable with the peak value  around 900~\AA\ (cf.~Figure~\ref{phota3pi}).  

 Moreover, in the inner cometary coma, below cometocentric distance of 50 km, the optical depth for  solar 
flux at wavelengths below 200~\AA\ and above 1000~\AA\ is smaller compared to  other 
wavelengths  because of smaller absorption cross sections of neutral species (mainly water). 
The rate of photodissociative excitation of CO$_2$ molecule into CO(\cam)  mainly depends 
on the degradation of solar flux 
in the wavelength region 850--1050 \AA. Hence, in  the innermost coma ($\le$50 km), for a 
given relative abundance of CO$_2$, the production rate of CO(\cam) via photodissociation of CO$_2$ is 
determined by the solar flux in the wavelength bin 1000--1050~\AA\ and  at wavelengths 
1025.7~\AA\ (H I) and 1031.9~\AA\ (O VI). 
 The calculated photodissociation rates of CO$_2$ 
producing CO(\cam) at 0.9 AU  are 1.66 $\times$ 10$^{-7}$ s$^{-1}$ and 5.28 $\times$ 10$^{-7}$ s$^{-1}$ 
using EUVAC and S2K solar fluxes, respectively, on 13 March 1986.

From Figure~\ref{csca3pi} it is seen that the  calculated  steady state photoelectron flux 
using two  solar flux models  differ in magnitude by a factor of 2. Since the cross section
for electron impact of CO producing CO(\cam)  peaks at lower energies ($\sim$10 eV) 
 where the photoelectron flux is also high ($\sim$10$^8$ cm$^{-2}$ s$^{-1}$ eV$^{-1}$  sr$^{-1}$; 
cf. Fig.~\ref{csca3pi}), the electron impact excitation of CO is a  major production  
source of Cameron band emission. At larger ($>$5000 km) cometocentric distances, due to 
decrease in photoelectron flux, the photodissociative excitation of CO$_2$ starts becoming 
an increasingly important process (cf. Fig.~\ref{proda3pi}). Contributions from dissociative recombination reactions 
and resonance fluorescence of CO are more than two orders of magnitude lower compared to 
major production processes.

\begin{figure}  
\begin{center}
\noindent\includegraphics[width=20pc,angle=0]{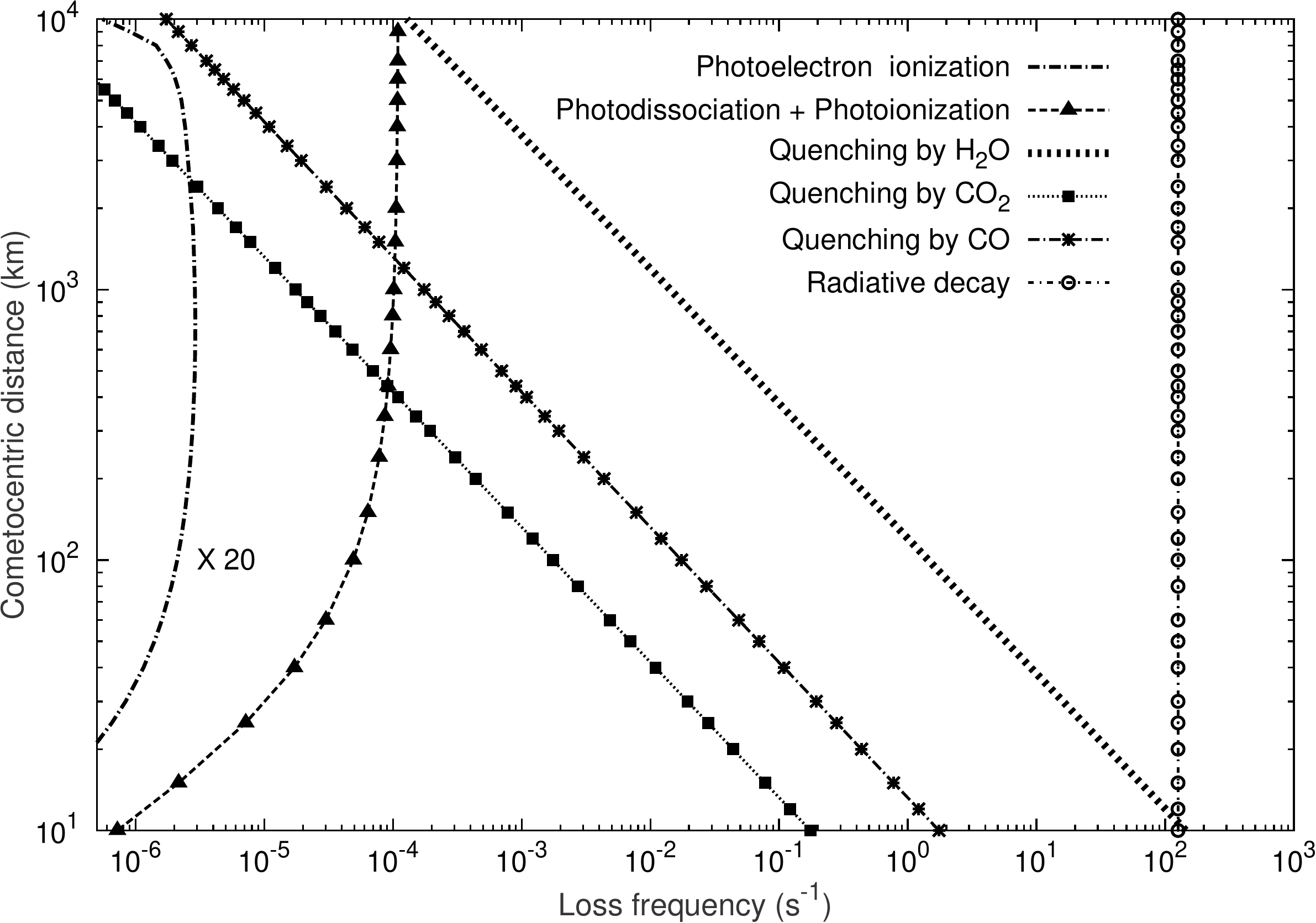}
\caption{Radial profiles of various loss mechanisms of CO(a$^3\Pi$) for 4\% CO$_2$ and 7\% CO relative 
abundances using EUVAC solar flux. Photoelectron impact ionization of 
CO(a$^3\Pi$) is plotted after multiplying by a factor 20.}
\label{lossa3pi}
\end{center}
\end{figure}

\begin{figure} 
\begin{center}
\noindent\includegraphics[width=20pc,angle=0]{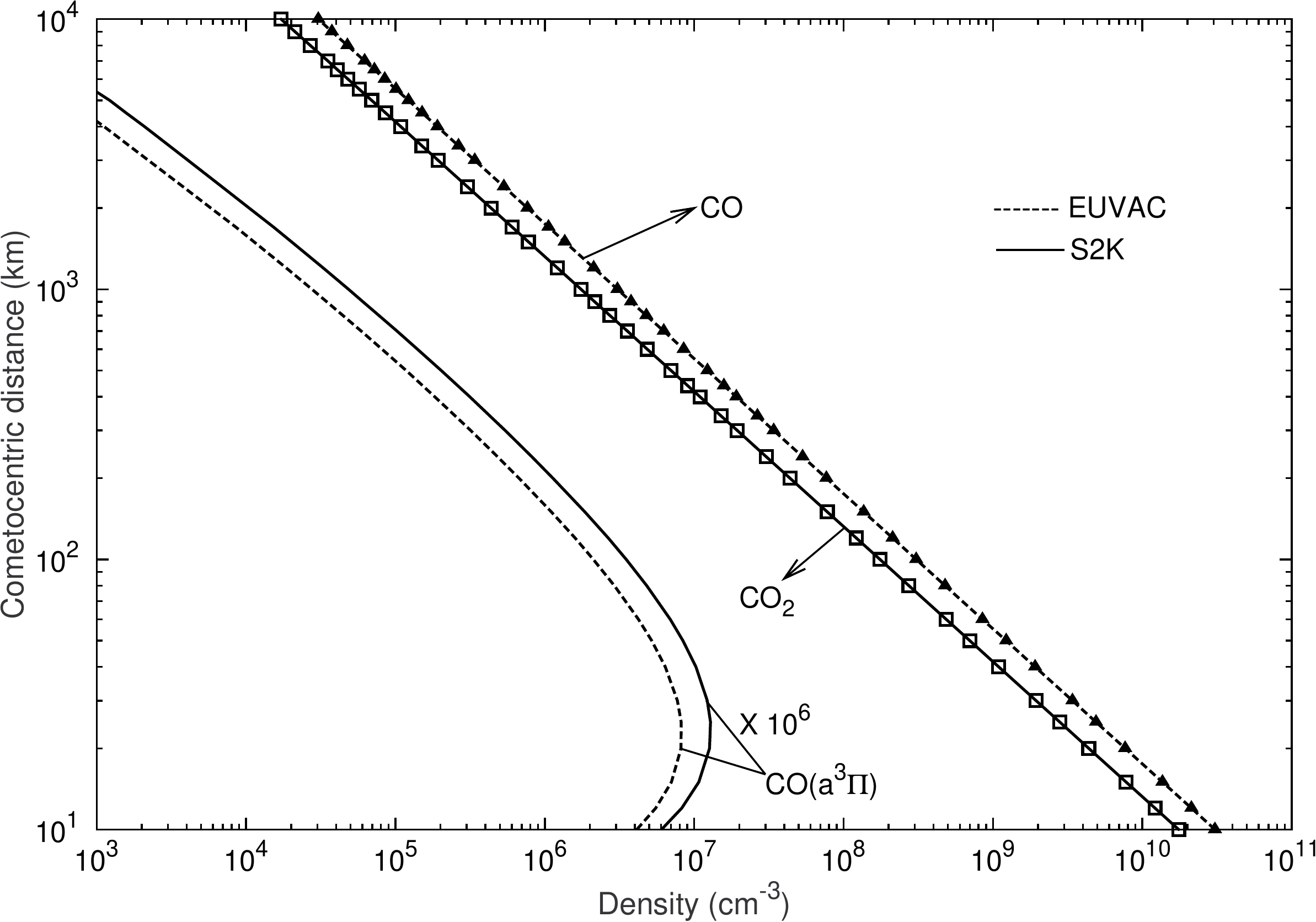}
\caption{The calculated radial profiles of number density CO(a$^3\Pi$) for SOLAR2000 (S2K) and EUVAC 
solar flux models. The density of CO(a$^3\Pi$) is plotted after multiplying  by a factor 10${^6}$.
 The number density profiles of CO$_2$ and CO are also shown for 4\% and 7\% relative abundances, 
respectively.}
\label{dena3pi}
\end{center}
\end{figure}

Since the lifetime of CO(a$^3\Pi$) is about
$\sim$3 ms, the quenching of the excited  \cam\ metastable state by various cometary
species is not very efficient. The calculated loss rate profiles of CO(\cam) for various 
processes  are shown in Figure~\ref{lossa3pi}. The radiative de-excitation is the main loss process.
Very close to the nucleus, the loss due to quenching of CO(a$^3\Pi$) by water is comparable to the 
radiative de-excitation. 
Quenching by water molecule would be a  more significant loss process of CO(\cam) in the  higher water 
production rate comets, like Hale-Bopp or when the comet is much closer to the Sun than 1 AU. 
The calculated  number density profile of CO(a$^3\Pi$) is shown in Fig.~\ref{dena3pi}.
 Above 100 km, the density profile of CO(a$^3\Pi$) mostly following the number density profiles 
of  the parent species CO$_2$ and CO. 

The above calculated total production rate is integrated up to 
10$^5$ km to obtain the height-integrated column intensity of Cameron band emission which is 
presented in Table~\ref{bigtaba3pi1}.
We  also calculated the line of sight intensity  at a given projected distance $z$ from the 
cometary nucleus using  production rates of different excitation processes of CO(\cam) as 
\begin{equation}
 I(z) = 2  \int_{z}^{R}V(s)\ ds
\end{equation}
where $s$ is abscissa along the line of sight and V($s$) is the corresponding emission rate.
 The maximum limit of integration $R$ is taken as 10$^5$ km.
 These brightness profiles are then averaged over the projected area 
 6600 $\times$ 11000 km$^2$ corresponding to the IUE slit dimension
  9$''$.07 $\times$ 15$''$.1 centred on nucleus of comet \halley\
 on 13 March 1986  at geocentric distance 0.96 AU. The volume emission rate for 3 transitions 
(0-0, 1-0, and 0-1) of the Cameron band are calculated using the following  formula 
\begin{equation}
V_{\nu'\nu''}(r)=q_{o\nu'} (A_{\nu'\nu''}/\sum_{\nu''}A_{\nu'\nu''}) V(r) exp(-\tau)
\end{equation}
where V(r) is total volume excitation rate of CO(a$^3\Pi$) at a given cometocentric distance $r$,
given by equation~\ref{volex},
q$_{o\nu'}$ is the Franck-Condon factor for transition, $A_{\nu'\nu''}$ is the Einstein 
transition probability from upper 
state $\nu'$  to lower state $\nu''$, and $\tau$ is the optical depth. Since resonance 
fluorescence is not an effective excitation mechanism for the Cameron band,
the cometary coma can be safely  assumed to be optically thin. The Franck-Condon 
factors are taken from \cite{Nicholls62} and branching ratios from \cite{Conway81}.

\begin{figure} 
\begin{center}
\noindent\includegraphics[width=20pc,angle=0]{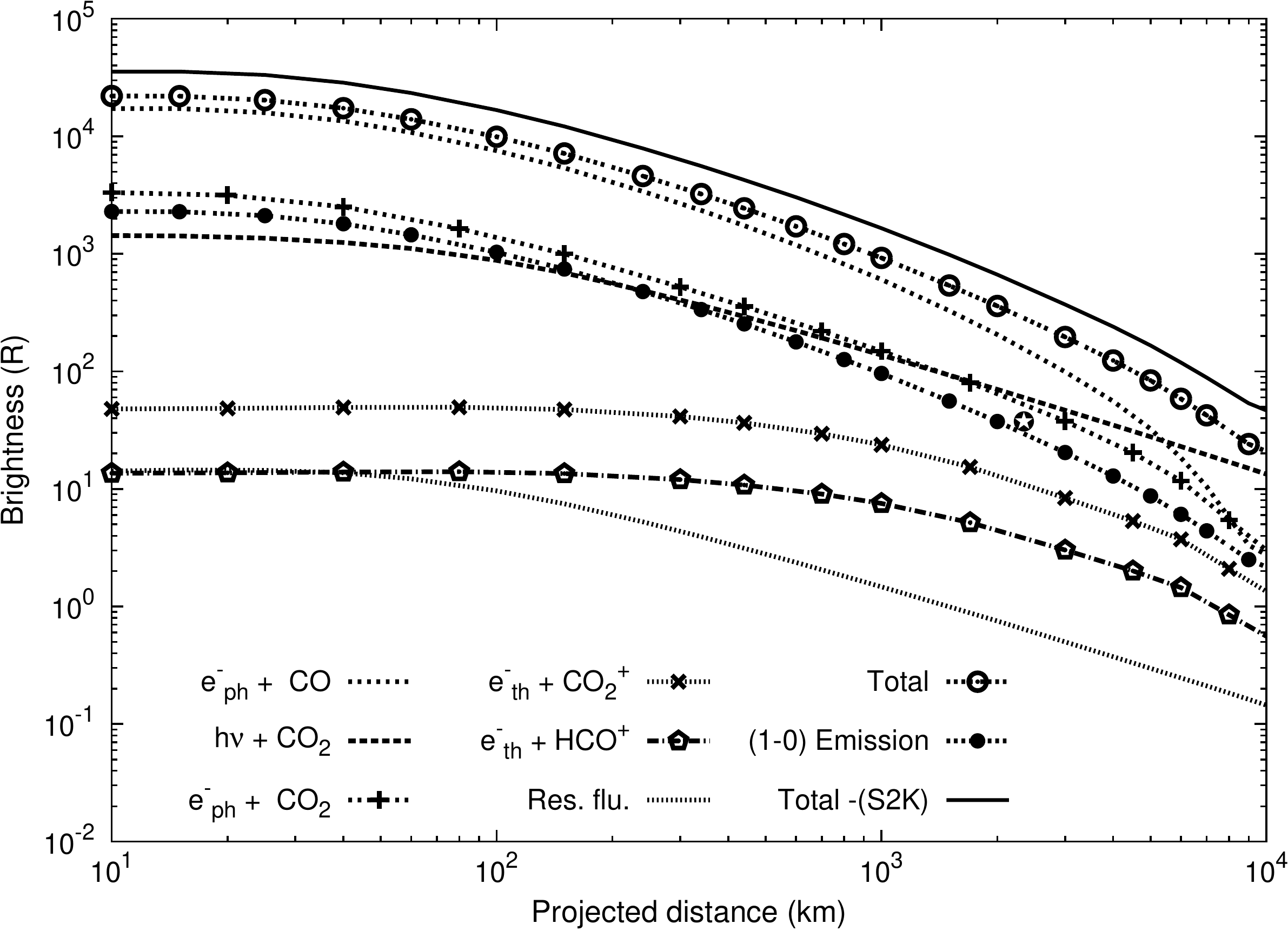}
\caption{The integrated Cameron band brightness profiles as a function of projected distance from nucleus
 for different production processes of the Cameron band, using EUVAC solar flux model and 
relative contribution of 4\% CO$_2$ and 7\%  CO. The calculated brightness profiles for Cameron (1-0)
band for EUVAC solar flux and total brightness for S2K solar flux are also shown.}
\label{proja3pi}
\end{center}
\end{figure}

The calculated brightness profiles for each of the production processes along projected 
distances from nucleus are shown in Figure~\ref{proja3pi}. 
At 100 km projected distance, the contribution due to photoelectron impact excitation 
of CO  to the total Cameron band intensity is about a factor 4 higher than  
the dissociative excitation processes of CO$_2$, 
while contributions of other production processes are around  2 orders of magnitude smaller.
 Around 1000 km projected distance,  both photodissociative excitation and electron impact 
dissociative excitation of CO$_2$  are contributing equally  to the total Cameron band 
intensity. The photodissociative 
excitation of CO$_2$ dominates the electron impact excitation processes above 5000 km.

The calculated relative contributions of (1-0), 
(0-0), and (0-1) bands to the total Cameron band are 13.9\%, 10.4\%, and 14.5\%, respectively.
The intensities of (1-0), (0-0) and (0-1) Cameron bands  of 
CO molecule are calculated as a function of relative abundances of CO$_2$ and CO.
The calculated percentage contributions of different production processes of
Cameron band  at three projected  distances for two  different solar flux models 
 are presented in Table~\ref{bigtaba3pi1}.  The IUE-observed 1-0 Cameron band 
emission on 13 March 1986 is 37$\pm$6 Rayleighs. 

\begin{figure} 
\begin{center}
\noindent\includegraphics[width=20pc,angle=0]{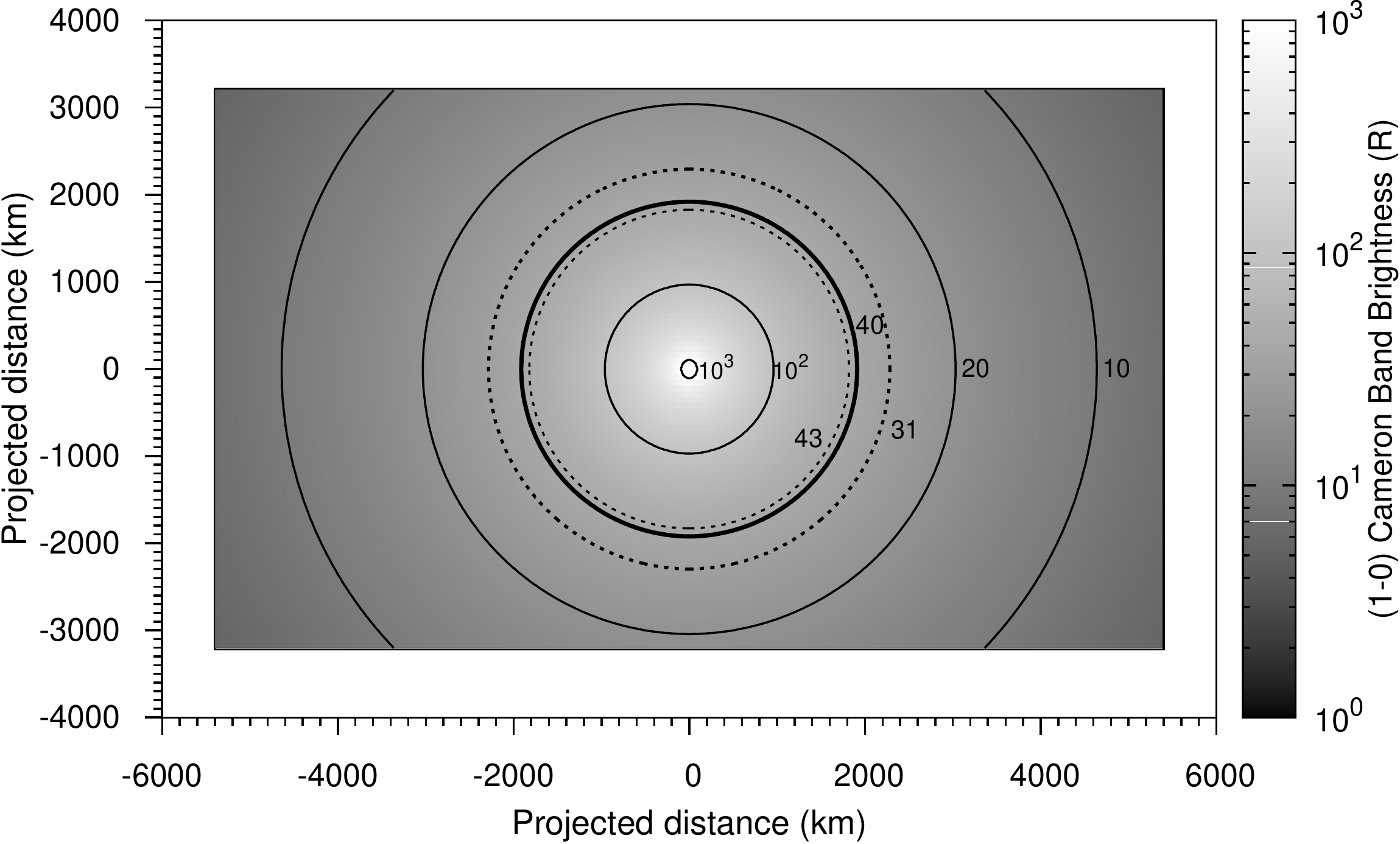}
\caption{The calculated (1-0) Cameron  band emission brightness in the IUE projected field of view 
on 13 March 1986, assuming spherical symmetry, using EUVAC solar flux model, for 
relative contribution of 4.3\% CO$_2$ and 4.7\% CO. The rectangle represent the 
 projected field of view corresponding  to IUE  
slit dimension of 9.07$''$ $\times$ 15.1$''$ centred on the nucleus of comet \halley, which is  
11000 $\times$ 6600 km.  The gray scale represent the calculated brightness with contours 
(solid lines) for 10$^3$, 10$^2$, 20, and 10 R marked in the figure. The calculated 
brightness  averaged over 
IUE slit projected area   (40 R) is shown by thick black contour between two dotted line contours which 
 represent the upper and lower limits of IUE observed intensity value (37 $\pm$ 6 R).}
\label{proj10}
\end{center}
\end{figure}

Using EUVAC solar flux as input, our model calculated 1-0 Cameron band emission intensity for the
relative abundance 4\% CO$_2$ and extended distribution of CO is 59 Rayleighs which is higher than 
IUE observed intensity by a factor 1.3 to 2. Taking CO$_2$ abundance as 4\% and CO abundance as 
 7\% from nucleus, the 
calculated 1-0 intensity is 51 Rayleighs, which is higher 
than the IUE-observed value by a factor 1.2 to 1.6.
The calculated intensity for 3\% CO$_2$ and 7\% CO  is 46 Rayleighs, which is 
consistent only  with the upper  limit of IUE-observed intensity. 
In all the above cases, below 1000 km projected distances, the contribution of 
photodissociation of CO$_2$ to the Cameron 
band emission is $<$15\%, while  electron impact of CO contribute 65 to 80\%. 
We have also calculated the intensity of Cameron band taking the \cite{Feldman97}
derived abundances of 4.3\% CO$_2$ and 4.7\% CO.
 The calculated intensity of 1-0 Cameron band emission in this case is 40 R, which is
 consistent with  the observed value of 37 $\pm$ 6 R
on 13 March 1986 (cf.~Table~\ref{bigtaba3pi2}). 
The calculated 1-0 Cameron band emission intensity at various projected distances 
in the IUE-slit field of view is presented in the Figure~\ref{proj10}; The circular contours
and gray scale provide information on brightness variation.
The calculated results using  S2K solar flux model for the above discussed relative 
compositions of CO$_2$ and CO are  also presented in Table~\ref{bigtaba3pi1}. The calculated  intensities 
are higher by a factor of $\sim$1.5, which is  mainly due to higher input solar flux 
and subsequently EUV-produced photoelectron's flux (cf.~Figs.~\ref{solflx}~and~\ref{csca3pi}).

 Using OH 3085~\AA\ emission observation by IUE, \cite{Tozzi98} derived water production 
rates for different days of IUE observations (1986 March 9, 11, 13, 16, 18) around Giotto encounter period.
 The water production rate derived on 13 March 1986, the closest approach day of Giotto 
spacecraft, was 5.9 $\times$ 10$^{29}$ s$^{-1}$.
 \cite{Feldman97} have considered these derived production rates of H$_2$O to estimate relative 
abundances of CO$_2$ and CO  for corresponding days of observation.
We  have calculated the intensity of Cameron band for different days of IUE observations taking 
the same H$_2$O, CO$_2$, and CO production rates as quoted in \cite{Feldman97}. The solar EUV fluxes 
on each day of observation was obtained by using EUVAC and S2K solar flux models and scaling them according 
to the heliocentric distance of comet. 
The IUE projected field of view is calculated for IUE slit dimension used in observation,
which  vary according to the geocentric distance of the comet in March 1986. 
The calculated intensities of Cameron 1-0, 0-0, 0-1 bands and percentage contributions from different 
production process to the  IUE slit-averaged brightness 
are presented Table~\ref{bigtaba3pi2}.
The calculated intensity of 1-0 emission is consistent with the IUE-observation for the EUVAC solar 
flux model,  while it is higher by a factor of 1.5 on using S2K solar flux. 
The calculations 
presented in Table~\ref{bigtaba3pi2} show that for a change in the CO$_2$/CO  abundance ratio
by a factor of 2, the total photoelectron impact excitation  contribution changes by only $\sim$10\%; it 
varies from  68 to 76\% (60 to 69\%) of the total IUE-observed intensity for EUVAC (S2K) solar flux model. 
The photoelectron impact excitation of CO alone contribute 
around 45 to 55\% (40 to 60\%)  to the total Cameron band intensity when EUVAC (S2K) solar flux is 
used. The contribution of photodissociation of CO$_2$ to the 
 IUE-observed   Cameron band 
brightness is around 20\% (30\%) for EUVAC (S2K) solar flux model when the abundances of CO and CO$_2$ in 
the comet are almost equal. These computation show that in the IUE field of view the 
 photoelectron is a major production  source (60-75\% contribution) for the Cameron band emission, 
whereas  the contribution due to photons is small (20-35\%).

The calculations  presented in Tables~\ref{bigtaba3pi1} and~\ref{bigtaba3pi2}  renders 
 that in case of comets where CO$_2$/CO 
abundance ratio is closer to 1 or larger than  1, the emission intensity of Cameron band is mainly 
controlled by the abundance of CO in the inner cometary coma. The photoelectron impact excitation 
of CO is the main production mechanism for the production 
of Cameron band emission, but not the photodissociative excitation of CO$_2$ as suggested or assumed in 
earlier studies \citep{Weaver94,Weaver97,Feldman97}.  Thus, in comets that have sufficient CO abundance  
the electron impact excitation of CO producing CO(\cam) can be an efficient excitation mechanism for 
Cameron band emission. Since Cameron emission is mainly governed by electron impact excitation reactions,
 this emission can be
used to track the photoelectron density mainly in the energy range 10 to 15 eV near the nucleus.

In the case of comet \hartley, which has an order of magnitude lower gas production rate  and
 much lower CO (abundance $<1$\%) than comet \halley, the dissociative recombination of CO$_2^+$ becomes
 a  competing production mechanism at larger ($>$10$^4$ km) cometocentric 
distances \citep{Bhardwaj11a}. 
However, in comparison, on comet \halley\ the production rates of  H$_2$O, CO$_2$, and CO are so high that 
the photon and photoelectron 
impact reactions are dominant throughout the inner cometary coma.
\subsection{Effect of electron impact cross section}
In this section we will discuss the impact of cross sections for electron impact excitation of CO(\cam)
 from CO$_2$ and CO. The threshold for  exciting CO molecule 
in the  metastable \cam\ state is 6 eV and the peak value of cross section occurs  around 
10 eV (cf.~Fig.~\ref{csca3pi}). The cross section for electron impact excitation of CO producing 
CO(\cam) reported 
by \cite{Jackman77} is theoretically fitted based on Born approximation and  experimental measurements
of \cite{Ajello71}. The uncertainty associated with  measurement is 
about 75\%. However, the uncertainty in the  cross section 
 at energies less than 15 
eV is 35\% \citep{Ajello71}, 
where the contribution of electron impact excitation plays a major role (cf. Fig~\ref{csca3pi}).
 The  cross section  measurements  of \cite{Furlong96} 
 differ at the peak value of cross section by a factor 2 
(cf.~Figure~\ref{csca3pi}). The threshold for dissociation of CO$_2$ 
molecule into CO(\cam)
 state is 11.45 eV. \cite{Ajello71} measured Cameron band emission cross sections in the wavelength region 
1950--2500~\AA\ by exciting  CO$_2$ molecule through electron impact. \cite{Sawada72} concluded that these
 cross sections are comparable with cross sections of 12.6 eV and 13.6 eV states. 
 The  cross section value for CO(\cam) production due to electron impact of CO$_2$
 measured at 80 eV by \cite{Erdman83} is 2.4 $\times$ 10$^{-16}$ cm$^{-2}$. 
\cite{Bhardwaj09} modified the fitting parameters given by \cite{Jackman77} for the 
excited states 12.6 eV and 13.6 eV of CO$_2$ molecule  to match cross section value measured 
by \cite{Erdman83} at 80 eV \citep[for more discussion on these cross sections 
see][]{Ajello71,Sawada72,Bhardwaj09}.
 \cite{Avakyan98} corrected \cite{Ajello71} reported cross sections based on measurements 
of \cite{Erdman83}. The difference in the  cross section of \cite{Avakyan98} and \cite{Bhardwaj09} 
   below 30 eV is about a factor of 2 (cf. Fig~\ref{csca3pi}). 

Using electron impact CO(\cam) excitation cross sections from \cite{Furlong96} for CO and from
 \cite{Avakyan98} for CO$_2$,  and using EUVAC solar flux, the calculated emission 
intensity of 1-0 Cameron band, for a given relative abundances of CO and CO$_2$, is larger 
by a factor 2. In these calculations the contribution of electron impact excitation of CO 
is increased from 70\% to 85\% at cometocentric distances below 10$^3$ km and  40\% to 60\%  
at distances above 10$^3$ km. On using these cross sections, 
the percentage  contribution  of photoelectron impact excitation  of CO to the total Cameron 
emission in the IUE-slit-averaged intensity  
is found to increase by 10\%, but there is no significant change 
in electron impact excitation of CO$_2$. In this case the contribution from photodissociative
 excitation of CO$_2$ is decreased by 10\%. 
 
 \section{ Summary}      
   Using the coupled chemistry-emission model  a detailed study of Cameron band 
(a$^3\Pi~\rightarrow$~X$^1\Sigma^+$) emission has been carried out on the comet \halley\ around 
the Giotto encounter period. The effects of change in solar flux on the production of CO(\cam)  and 
thus the Cameron band intensity have been evaluated by considering two different solar EUV flux models, 
viz. EUVAC model \citep{Richards94} and S2K (SOLAR2000) model \citep{Tobiska04}. 
Calculations are made for different days of IUE-observation of comet \halley. 
 The important results from the present model calculations can be summarized as follows: 
\begin{itemize}

\item For the same day, the solar flux  from the two models (EUVAC \& S2K) are different, and the 
difference between them varies with wavelength.

\item The production rates obtained by using S2K solar flux model are higher than that of EUVAC model. 
The photodissociation of CO$_2$ is larger by a factor of 2.5, while the photoelectron impact excitation 
is larger by a factor of $\sim$1.5.

\item The total production rate of CO(\cam)  peaks around cometocentric distance of 20 km  
for both  solar flux models. 

\item Throughout the inner coma the main  loss mechanism of CO(\cam)  is 
radiative decay. Very close to the nucleus ($<$ 20 km)  quenching by water is also significant.

\item In the inner ($\le$ 5000 km) coma the major production mechanism of CO(\cam) is photoelectron 
impact excitation of CO.

\item On using EUVAC solar flux, and abundances of CO and CO$_2$ as derived from IUE-observation, 
 the model calculated  Cameron band  1-0 emission intensity (40 R) is consistent with the IUE-observed  
 brightness (37 $\pm$ 6 R)  on 13 March 1986, and also 
on other days of observations. However, the calculated intensities are larger by a factor 1.5 when the 
 S2K solar EUV flux is used.

\item For EUVAC (S2K) solar flux model, around 70\% (65\%) of the total intensity of Cameron band observed 
by the IUE is contributed by electron impact excitation of CO and CO$_2$ molecules,  while 
the contribution from photodissociative 
excitation of CO$_2$ is about 20--30\% only.
 
\item  In  comets having comparable CO and CO$_2$ relative abundances, 
the intensity of Cameron band is largely determined by the photoelectron
impact excitation  of CO, and not the photodissociative excitation of CO$_2$ as 
suggested by earlier studies.

 \item Since the emission intensity of  Cameron band is mainly governed by electron 
impact reactions, this emission may be more useful  to track the photoelectron density 
in 10--15 eV energy region  in the inner coma, rather than the CO$_2$ abundance. 
 \end{itemize}

\section*{Acknowledgements}
One of the authors (SR) was supported by ISRO research fellowship during the period of this study.

\clearpage





\renewcommand{\thefootnote}{\fnsymbol{footnote}}

 \begin{center}    
\begin{table} 
\caption{Reactions for the production and loss of CO($a^3\Pi$).}
\begin{tabular}{lll}
\hline
\multicolumn{1}{c}{ Reaction}
&\multicolumn{1}{l}{ Rate(cm$^{3}$ s$^{-1}$ or s$^{-1}$)}
&\multicolumn{1}{c}{ Reference} \\
\hline
CO$_2$ + h$\nu$ $\rightarrow$ CO(a$^3\Pi$) + O($^3$P)
& Model & \textit{Present work}\\
CO + h$\nu$ $\rightarrow$ CO(a$^3\Pi$)
&1.69$\times$ 10$^{-9}$ &\cite{Weaver94}\\
CO$_2$ + e$^-_{ph}$ $\rightarrow$ CO(a$^3\Pi$) + O + e$^-$
&Model&\textit{Present work}\\
CO + e$^-_{ph}$  $\rightarrow$ CO(a$^3\Pi$) + e$^-$ 
 &Model &\textit{Present work}\\
CO$_2^+$  + e$^-$ $\rightarrow$ CO(a$^3\Pi$) + O  
&{K$_a$}\footnotemark[1] &\cite{Seiersen03},\\
& & \cite{Rosati03}\\
HCO$^+$ + e$^-$ $\rightarrow$ CO(a$^3\Pi$) + H
&{K$_b$}\footnotemark[2]
&\cite{Rosati07},\\
& & \cite{Schmidt88}\\
CO(a$^3\Pi$) + h$\nu$ $\rightarrow$ C + O
&7.2$\times$ 10$^{-5}$  & \cite{Huebner92} \\
CO(a$^3\Pi$) + h$\nu$ $\rightarrow$ CO$^+$ + e$^-$
&8.58$\times$ 10$^{-6}$  & \cite{Huebner92} \\
CO(a$^3\Pi$) + h$\nu$ $\rightarrow$ O+C$^+$ + e$^-$
&2.45$\times$ 10$^{-8}$  & \cite{Huebner92} \\
CO(a$^3\Pi$) + h$\nu$ $\rightarrow$ C + O$^+$ + e$^-$
&2.06$\times$ 10$^{-8}$  & \cite{Huebner92}\\
CO(a$^3\Pi$) +H$_2$O $\rightarrow$ CO + H$_2$O  &3.3 $\times$ 10$^{-10}$  
& \cite{Wysong00}\\
CO(a$^3\Pi$) + CO$_2$ $\rightarrow$ CO + CO$_2$  &1.0 $\times$ 10$^{-11}$  
&\cite{Skrzypkowski98} \\
CO(a$^3\Pi$) + CO $\rightarrow$ CO + CO  &5.7 $\times$ 10$^{-11}$ 
&\cite{Wysong00}\\
CO(a$^3\Pi$) + e$^{-}_{ph}$ $\rightarrow$ CO$^{+}$ + 2e$^{-}$  & Model
&\textit{Present work}\\
 CO(a$^3\Pi$)\hspace{0.3cm}  $\longrightarrow$ CO  +  h$\nu$ & 1.26 $\times$ 10$^{2}$  
& \cite{Lawrence72} \\
\hline \\
\end{tabular}
\addtocounter{footnote}{-2}

\footnotemark[1]{\small K$_a$ =  6.5 $\times$ 10$^{-7}$ (300/Te)$^{0.8}$ $\times$ 
0.87 $\times$ 0.29 cm$^{3}$ s$^{-1}$; here 0.87 is yield of dissociative recombination 
of CO$_2^+$ producing CO, and 0.29 is yield of CO(\cam) produced from  CO.}

\footnotemark[2]{\small K$_b$ = 2.4 $\times$ 10$^{-7}$ (300/Te)$^{0.7}$ $\times$ 0.23 cm$^{3}$ s$^{-1}$;
here 0.23 is yield of dissociative recombination of HCO$^+$ producing CO(a$^3\Pi$), e$^-_{ph}$ = photoelectron. }
  \label{taba3pi}
\end{table}
\end{center}

\clearpage

\begin{center}     
\begin{sidewaystable} 
\small
\caption{{ Calculated brightness of the Cameron band at comet 
1P/Halley  for different conditions on 13 March 1986.}}
\label{bigtaba3pi1}
\scalebox{1}{
 \begin{tabular}{|c|c|c|c|c|c|c|c|c|c|c|c|c|c|c|c|c|c|c|c|c|}
\hline
\multicolumn{2}{|p{0.8 in}|}{\multirow{1}{0.8 in}{\centering Relative abundance}}
&\multicolumn{3}{m{1 in}|}{\multirow{4}{1.4 in}{\centering IUE-slit averaged  brightness (R)}}

&\multicolumn{12}{p{4.5 in}|}{\multirow{2}{4.5 in}{\centering {Percentage contribution to 
total Cameron band for different processes at three 
different projected radial distances (km)}} }
&\multicolumn{2}{p{1.8 in}|}{\centering Total Cameron band brightness (R)}
\\
\cline{1-2} \cline{6-19}
 \multicolumn{1}{|p{0.3 in}|} { \centering CO$_2$ (\%)} 
&\multicolumn{1}{p{0.2 in}|}{ \centering CO (\%)} 
&\multicolumn{3}{c| }{ }  
&\multicolumn{3}{p{1 in}|}{\multirow{3}{1 in} {\centering {h$\nu$ + CO$_2$} }}  
&\multicolumn{3}{p{1 in}|}{\multirow{3}{1 in} {\centering {e$^-_{ph}$ + CO$_2$}}}    
&\multicolumn{3}{p{1 in}|}{\multirow{3}{1 in} {\centering {e$^-_{ph}$ + CO}}}   
&\multicolumn{3}{p{1 in}|}{\multirow{3}{1 in} {\centering {e$^-$ + CO$_2^+$}} }
&\multicolumn{1}{p{0.7 in}|}{\centering IUE-slit averaged}  & 
\multicolumn{1}{p{1 in}|}{\centering Height integrated column}\\ 
\hline
 &  &\multicolumn{1}{c|}{(1-0)\footnotemark[1]} &\multicolumn{1}{c|}{(0-0)} 
&\multicolumn{1}{c|}{(0-1)}
&\multicolumn{1}{p{0.2 in}|}{10$^2$ }
&\multicolumn{1}{p{0.3 in}|}{10$^3$ }
&\multicolumn{1}{p{0.1 in}|}{10$^4$ }
&\multicolumn{1}{p{0.2 in}|}{10$^2$ } 
&\multicolumn{1}{p{0.3 in}|} {10$^3$ }
&\multicolumn{1}{p{0.1 in}|} {10$^4$ }
&\multicolumn{1}{p{0.2 in}|} {10$^2$ } 
&\multicolumn{1}{p{0.3 in}|} {10$^3$ }
&\multicolumn{1}{p{0.1 in}|} {10$^4$ }
&\multicolumn{1}{p{0.2 in}|} {10$^2$ } 
&\multicolumn{1}{p{0.3 in}|} {10$^3$ }
&\multicolumn{1}{p{0.1 in}|} {10$^4$ }
& & \\  
 \cline{3-17}
 \multicolumn{2}{|l|}{ \centering EUVAC} 
    &     &   &  &&  &   &    &     &  &   &   &    &  &   & & &  \\
     4 & Ext\footnotemark[2] & 59 & 44 & 63 & 9 & 14 & 52 & 15 & 15 & 11 & 74 & 66 &  
25 & 0.5 & 2 & 5 & 430 & 10946 \\ 
 4 & 7 & 51 & 38 & 54 & 9 & 15 & 65 & 14 & 16 & 13 & 75 & 64 & 12 
& 0.5 & 3 & 5 & 308 & 8836 \\
 3 & 7 & 46 & 34 & 48 & 7 & 12 & 63 & 11 & 13 & 12 & 80 & 70 & 15 
& 0.5 & 2 & 5  & 331  & 10626  \\
 4.3 & 4.7 & 45 & 34 & 48 & 11 & 19 & 69 & 20 & 21 & 13 & 68 &  
55 & 9 & 0.5 & 3 & 6 & 329 & 9582 \\

\multicolumn{2}{|l|}{\centering {S2K}} 
&  &        &  &&  &   &    &     &  &   &   &    &  &   & & &  \\
4  & Ext & 87 & 66 & 96 & 14 & 19 & 61 & 13 & 13 & 9 & 71 & 62 &  
20 & 0.5 & 3 & 4 & 638 & 15612 \\
     4 & 7 & 77 & 58 & 82 & 14 & 21 & 73 & 13 & 14 & 10 & 71 & 60 & 10 & 0.5 & 
2 & 5  & 559 & 15841 \\
 3 & 7 & 68 & 51 & 72 & 11 & 17 & 71 & 10  & 12 & 9 & 77 & 66 & 11  
& 0.5 & 2 & 5  & 490  & 14991 \\
 4.3 & 4.7 & 67 & 50 & 70 & 16 & 24 & 76 & 17 & 18 & 10 & 65 & 52 &   
6 & 0.5 & 3 & 5 & 472 & 13582 \\
 \hline
\end{tabular}}
\addtocounter{footnote}{-3}

\footnotemark[1]{\small{The intensity of Cameron  (1-0) band observed by IUE is  37$\pm$6
Raleighs on 13 March 1986.;}}
\footnotemark[2]{\small{Ext: Extended CO distribution.;
  e$^-_{ph}$ is photoelectron and e$^-$ is thermal electron.}};
\end{sidewaystable}
 \end{center}

\clearpage

\begin{center} 
\begin{sidewaystable}   
\small
\caption{{Calculated brightness of the  Cameron band at comet 
1P/Halley  on different days of IUE observations.}}
\label{bigtaba3pi2}
\scalebox{1}{
 \begin{tabular}{|c|c|c|c|c|c|c|p{0.8 in}|c|c|c|c|c|c|c|c|c|c|c|c|c|}
\hline
\multicolumn{1}{|m{0.5 in}|}{\multirow{4}{0.5 in}{\centering  Date in March 1986}} &
\multicolumn{1}{m{0.26 in}|}{\multirow{4}{0.3 in}{\centering  r (AU)}} &
\multicolumn{1}{m{0.26 in}|}{\multirow{4}{0.3 in}{\centering  $\Delta$ (AU)}} &
\multicolumn{1}{m{0.3 in}|}{\multirow{4}{0.3 in}{\centering Q$_{H_2O}$ (10$^{29}$ s$^{-1}$)\footnotemark[2] }} &
\multicolumn{2}{m{0.8 in}|}{\multirow{3}{0.9 in}{\centering  {Derived abundances (\%)}\footnotemark[2]}} &
\multicolumn{1}{m{0.4 in}|}{\multirow{4}{0.2 in}{\centering Ratio Q$_{CO_2}$ /Q$_{CO}$}} 
&\multicolumn{3}{m{1 in}|}{\multirow{3}{1.5 in}{\centering IUE-slit averaged brightness (R)  }}
&\multicolumn{4}{p{3.2 in}|}{{\centering {Percentage contribution to the IUE-slit averaged 
total Cameron band emission for different excitation processes (\%)}}} 
&\multicolumn{1}{p{0.8 in}|}{\multirow{1}{0.8 in}{\centering  {IUE-slit averaged~total brightness (R) }}}
\\  
\cline{5-6}  \cline{8-14}
 & & &  & CO$_2$ & CO &&
\multicolumn{1}{c|}{\centering {(1-0)}} &\multicolumn{1}{c|}{(0-0)} 
&\multicolumn{1}{c|}{(0-1)}
&\multicolumn{1}{p{0.67 in}|} {\centering {h$\nu$ + CO$_2$}  }   
&\multicolumn{1}{p{0.71 in}|} {\centering {e$^-_{ph}$ + CO$_2$} }     
&\multicolumn{1}{p{0.7 in}|} {\centering {e$^-_{ph}$ + CO}  }   
&\multicolumn{1}{p{0.7 in}|} {\centering {e$^-$ + CO$_2^+$} }
&  \\    
\hline
  
 \multicolumn{1}{|m{0.5 in}|}{ EUVAC} 
   &  &   &   &  &   &   &   &   &&& && &       \\
   9 &0.84&1.07& 7.50& 6.0& 6.5 & 0.92& 75 [64 $\pm$ 9]\footnotemark[1] & 57 & 80 & 22 & 22  & 51  &  3  & 550 \\
   11&0.87&1.02& 5.84& 5.1& 4.3 & 1.2 & 43 [43 $\pm$ 8]& 32 & 45 & 25 & 24  & 44  & 4  & 310  \\
   13&0.90&0.96& 5.98& 4.3& 4.7 & 0.9 & 40 [37 $\pm$ 6]& 30 & 43 & 22 & 21  & 51  & 4  &  293 \\
   16&0.95&0.89& 4.90& 6.3& 8.2 & 0.77& 42 [44 $\pm$ 9]& 32 & 45 & 23 & 20  & 51  & 4  & 307  \\
   18&0.97&0.84& 4.92& 2.8& 4.1 & 0.68& 24 [20 $\pm$ 6]& 18 & 26 & 20 & 19  & 57  & 3  & 177  \\

 \multicolumn{1}{|l|}{ S2K} 
  &&&&&   &  &   &   &  &   &   &   &    &       \\
   9  &0.84&1.07& 7.50 & 6.0& 6.5 & 0.92& 116 [64 $\pm$ 9] & 87 & 123 & 31 & 19  & 44  & 4 & 837 \\
   11 &0.87&1.02& 5.84& 5.1& 4.3  & 1.2 &  66 [43 $\pm$ 8] & 49 & 69  & 33 & 21  & 39  & 4 & 475  \\
   13 &0.90&0.96& 5.98& 4.3& 4.7  & 0.9 &  62 [37 $\pm$ 6] & 46 & 65  & 30 & 19  & 45  & 4 & 446 \\
   16 &0.95&0.89& 4.90& 6.3& 8.2  & 0.77&  64 [44 $\pm$ 9] & 48 & 68  & 30 & 18  & 45  & 4 & 456 \\
   18 &0.97&0.84& 4.92& 2.8& 4.1  & 0.68&  37 [20 $\pm$ 6] & 28 & 39  & 25 & 16  & 53  & 3 & 262 \\

 \hline
\end{tabular}}
\addtocounter{footnote}{-3}

\footnotemark[1]{\small{The value in square brackets is IUE-observed (1-0) Cameron band intensity;
e$^-_{ph}$ = photoelectron, e$^-$ = thermal electron.}}
\footnotemark[2]{\small{The production rates of H$_2$O  and abundances of CO$_2$ \& CO
 are taken from  \cite{Feldman97}.}}

\end{sidewaystable}
 \end{center}
\newpage


\end{document}